\documentstyle[12pt,epsfig]{article}
\begin{document}
\def\scri{\unitlength=1.00mm
\thinlines
\begin{picture}(3.5,2.5)(3,3.8)
\put(4.9,5.12){\makebox(0,0)[cc]{$\cal J$}}
\bezier{20}(6.27,5.87)(3.93,4.60)(4.23,5.73)
\end{picture}}

\noindent Stockholm\\
USITP 05-4\\
September 2005\\
Revised November 2005\\

\vspace{1cm}

\begin{center}

{\Large ANTI DE SITTER SPACE,}

\vspace{5mm}

{\Large SQUASHED AND STRETCHED}

\vspace{1cm}

{\large Ingemar Bengtsson}\footnote{Email address: ingemar@physto.se. 
Supported by VR.}

\

{\large Patrik Sandin}

\vspace{1cm}

{\sl Stockholm University, AlbaNova\\
Fysikum\\
S-106 91 Stockholm, Sweden}

\vspace{8mm}

{\bf Abstract}

\end{center}

\vspace{5mm}

\noindent We study the Lorentzian analogues of the squashed 3-sphere, 
namely 2+1 dimensional anti-de Sitter space squashed or stretched 
along fibres that are either spacelike or timelike. The causal 
structure, and the property of being an Einstein--Weyl space, 
depend critically on whether we squash or stretch. We argue that 
squashing, and stretching, completely destroys the conformal 
boundary of the unsquashed spacetime. As a physical application 
we observe that the near horizon geometry of the extremal Kerr 
black hole, at constant Boyer--Lindquist latitude, 
is anti-de Sitter space squashed along compactified spacelike fibres.  

\newpage

{\bf 1. Introduction}

\vspace{5mm}

\noindent The Hopf fibration of the 3-sphere appears throughout mathematical 
physics in many guises; it is used to describe qubits, magnetic monopoles, 
Taub-NUT universes, and what not. There is a beautiful picture behind it: 
the Hopf fibres form a space-filling congruence of linked geodesic 
circles in the 3-sphere. In the Taub-NUT cosmologies the 3-sphere 
is squashed along the Hopf fibres. Such spheres 
are known as Berger spheres by mathematicians. They are solutions to 
the conformally invariant Einstein--Weyl equations.

The squashed 3-sphere has a Lorentzian analogue. In fact it has two 
Lorentzian analogues, since 3 dimensional anti-de Sitter space ${\bf adS}_3$ 
can be squashed (or stretched) along Hopf fibres that are either spacelike 
or timelike. This construction was briefly discussed by Jones, Tod and 
Pedersen \cite{Jones} \cite{Pedersen}, because such spacetimes admit a 
twistorial description (with a two dimensional family of totally geodesic 
null hypersurfaces serving as twistor space \cite{Hitchin}). From this point 
of view squashed anti-de Sitter space becomes interesting as a simple but 
non-trivial example in twistor theory. It has also been studied as an 
asymmetric deformation of the conformal field theory that describes the 
propagation of strings on the group manifold of $SL(2, {\bf R})$---also known 
as ${\bf adS}_3$ \cite{Israel, Detournay}. But there are other uses of such 
a natural construction, in particular the near horizon geometry of the 
extremal Kerr black hole \cite{Gary} can be understood using it. For 
this reason we have studied squashed anti-de Sitter space in some detail. 
We also use it to 
point a moral: we will argue that the squashing completely destroys the 
conformal boundary of the unsquashed spacetime. This tells us that conformal 
compactification \cite{Penrose} depends much more on the detailed structure 
of Einstein's equations than one might perhaps think it would.  

The contents of this paper: We describe some relevant features of 2+1 
dimensional anti-de Sitter space in section 2, but since this has been 
described at length elsewhere---we recommend ref. \cite{Holst} and 
references therein---some details are relegated to an Appendix. In 
section 2 we concentrate on the two geodetic congruences, one 
timelike and one spacelike, that will play the role that 
the Hopf circles play for the 3-sphere. In section 3 we squash and 
stretch our spacetime along these fibres, discuss the symmetries of 
the resulting spacetimes, and find the Killing horizons that they 
contain. Section 4 makes some observations on null geodesics; the 
distinction between squashing and stretching now begins to become 
apparent. For timelike stretching detailed results are available 
already---we are in effect studying the G\"odel spacetime \cite{Spindel}. 
In section 5 we establish when our spacetimes solve the 
conformally invariant Einstein--Weyl equations. In section 6 we attempt 
to conformally compactify our spacetimes, and argue that the boundary is 
destroyed by squashing (and stretching). Section 7 applies what we 
have learned to a discussion of the extremal Kerr black hole. 
Conclusions and open questions are briefly listed in section 8.

\vspace{1cm}

{\bf 2. Geodetic congruences in anti-de Sitter space}

\vspace{5mm}

\noindent Anti-de Sitter space is defined as a quadric surface embedded 
in a flat space of signature $(+ \dots +--)$. Thus 2+1 dimensional 
anti-de Sitter space is defined as the hypersurface   

\begin{equation} X^2 + Y^2 - U^2 - V^2 = - 1 \end{equation}

\noindent embedded in a 4 dimensional flat space with the metric 

\begin{equation} ds^2 = dX^2 + dY^2 - dU^2 - dV^2 \ . \end{equation}

\noindent The Killing vectors are denoted $J_{XY} = X\partial_Y - Y\partial_X$, 
$J_{XU} = X\partial_U + U\partial_X$, and so on. The topology is now 
${\bf R}^2 \times {\bf S}^1$, and one may wish to go to the covering 
space in order to remove the closed timelike curves. Our arguments 
will mostly not depend on whether this final step is taken. 

For the 2+1 dimensional case the definition can be reformulated in an 
interesting way. Anti-de Sitter space 
can be regarded as the group manifold of $SL(2,{\bf R})$, that is as the 
set of matrices

\begin{equation} g = \left[ \begin{array}{cc} V+X & Y+U \\ Y-U & V-X \end{array} 
\right] \ , \hspace{10mm} \mbox{det}g = U^2 + V^2 - X^2 - Y^2 = 1 
\ . \label{A3} \end{equation}

\noindent The group manifold is equipped with its natural metric, which is 
invariant under transformations $g \rightarrow g_1gg_2^{-1}$, $g_1, g_2 \in 
SL(2, {\bf R})$. The Killing vectors can now be organized into two orthonormal 
and mutually commuting sets, 

\begin{eqnarray} & J_1 = - J_{XU} - J_{YV} \hspace{15mm} 
& \tilde{J}_1 =  - J_{XU} + J_{YV} \\
& J_2 = - J_{XV} + J_{YU} \hspace{15mm} 
& \tilde{J}_2 = - J_{XV} - J_{YU} \\
& J_0 = - J_{XY} - J_{UV} \hspace{15mm} 
& \tilde{J}_0 = J_{XY} - J_{UV} \ . \end{eqnarray} 

\noindent They obey 

\begin{equation} ||J_1||^2 = ||J_2||^2 = - ||J_0||^2 = 1 \ , \hspace{3mm} 
||\tilde{J}_1||^2 = ||\tilde{J}_2||^2 = - ||\tilde{J}_0||^2 = 1 \ . \end{equation}

\noindent Locally $SL(2, {\bf R})$ is isomorphic with the Lorentz group $SO(2,1)$. 
The isometry group $SO(2,2)$ is therefore locally isomorphic to 
$SO(2,1)\times SO(2,1)$. These matters are discussed more fully in ref. 
\cite{Holst}. Very similar things can be said about the 3-sphere.

Here we would like to describe a coordinate system $(\tau, \omega, \sigma)$ 
\cite{Marc}, 
analogous to the Euler angles that are used to describe the 3-sphere. To this 
end we parametrize an arbitrary $SL(2, {\bf R})$ matrix as 

\begin{equation} g(\tau , \omega , \sigma ) = \left[ \begin{array}{cc} 
\cos{\frac{\tau}{2}} & - \sin{\frac{\tau}{2}} \\ 
\sin{\frac{\tau}{2}} & \cos{\frac{\tau}{2}} \end{array} \right] 
\left[ \begin{array}{cc} 
\sinh{\frac{\omega}{2}} & \cosh{\frac{\omega}{2}} \\ 
- \cosh{\frac{\omega}{2}} & - \sinh{\frac{\omega}{2}} \end{array} \right] 
\left[ \begin{array}{cc} 
\exp{-\frac{\sigma}{2}} & 0 \\ 0 & \exp{\frac{\sigma}{2}} \end{array} \right] 
\ . \end{equation}

\noindent Straightforward calculations show that the Killing vectors in 
the first $SO(2,1)$ factor are  

\begin{eqnarray} J_1 & = & -\frac{2\sinh{\sigma}}{\cosh{\omega}}\partial_\tau - 
2\cosh{\sigma}\partial_\omega + 2\tanh{\omega}\sinh{\sigma}\partial_\sigma 
\label{1} \\
J_2 & = & 2\partial_\sigma \label{2} \\
J_0 & = & \frac{2\cosh{\sigma}}{\cosh{\omega}}\partial_\tau + 
2\sinh{\sigma}\partial_\omega - 2\tanh{\omega}\cosh{\sigma}\partial_\sigma 
\ . \label{3} \end{eqnarray}

\noindent The second $SO(2,1)$ factor is spanned by

\begin{eqnarray} \tilde{J}_1 & = & 2\sin{\tau}\tanh{\omega}\partial_\tau - 
2\cos{\tau}\partial_\omega + \frac{2\sin{\tau}}{\cosh{\omega}}\partial_\sigma 
\label{4} \\
\tilde{J}_2 & = & -2\cos{\tau}\tanh{\omega}\partial_\tau - 
2\sin{\tau}\partial_\omega -\frac{2\cos{\tau}}{\cosh{\omega}}\partial_\sigma 
\label{5} \\
\tilde{J}_0 & = & 2\partial_\tau \ . \label{6} \end{eqnarray}

\noindent We will focus on the mutually commuting Killing vectors $J_2$ and 
$\tilde{J}_0$, to which our coordinate system is adapted. They form two 
nowhere vanishing vector fields in ${\bf adS}_3$. In any odd dimensional 
anti-de Sitter space we can construct a nowhere vanishing timelike vector 
field analogous to $\tilde{J}_0$, while there is no similar higher 
dimensional analogue for $J_2$. But in dimension 3 we have these two 
everywhere vanishing vector fields to play with. Each of them defines an 
interesting congruence in anti-de Sitter space, and their flow lines 
are the Hopf fibres along which we will squash and stretch our spacetime. 

The metric on anti-de Sitter space takes the form 

\begin{eqnarray} ds^2 & = & \frac{1}{4}( - d\tau^2 + d\omega^2 + d\sigma^2 
+ 2\sinh{\omega}d\tau d\sigma ) = \\
\ \nonumber \\ 
& = & \frac{1}{4}\left( - (d\tau - \sinh{\omega}d\sigma )^2 + 
d\omega^2 + \cosh^2{\omega}d\sigma^2 \right) \\
\ \nonumber \\
& = & \frac{1}{4}\left( - \cosh^2{\omega}d\tau^2 + 
d\omega^2 + (d\sigma + \sinh{\omega}d\tau )^2 \right) \ . \end{eqnarray}   

\noindent The flow lines 
of our two Killing vector fields are geodesics, so we are dealing with
two geodetic congruences. 

We will soon draw pictures of these congruences. To do so it is 
convenient to use another coordinate system, namely the sausage coordinates 
described in the Appendix. Then anti-de Sitter space itself will appear 
as a cylinder sliced with Poincar\'e disks of constant negative curvature.
 
One more remark about the symmetries of anti-de Sitter space is needed. 
(We make it brief, because it was fully spelt out elsewhere 
\cite{Holst,Peldan}.) 
We will be especially interested in the Killing horizons that arise. 
For what conjugacy classes of isometries does this happen? To answer 
this question one begins with the observation that the conjugacy classes 
of $SO(2,1)$ can be divided into hyperbolic, elliptic and parabolic 
transformations. Since the group manifold of $SO(2,1)$, or more precisely 
its double covering $SL(2, {\bf R})$, is itself a copy of ${\bf adS}_3$, 
these conjugacy classes correspond to two dimensional surfaces in the 
group manifold (with the parabolic conjugacy classes forming the forwards 
and backwards lightcones of the origin). Since the Lie algebra of $SO(2,2)$ 
is a direct product of two copies of the Lie algebra of $SO(2,1)$, it is 
then straightforward to divide the Killing vectors of the former group 
into conjugacy classes. Further, it is known that bifurcate 
Killing horizons in ${\bf adS}_3$ occur for conjugacy classes where 
the transformations take the form (hyperbolic) $\times $ (hyperbolic); 
they are numerous 
enough so that every spacelike geodesic is the bifurcation line of such 
a Killing horizon. Degenerate Killing horizons occur for transformations 
of the form (parabolic) $\times $ (parabolic). They form a two parameter 
family of totally geodesic null surfaces, and can be regarded as light 
cones with vertices on the conformal boundary \scri . Finally transformations 
of the type (parabolic) $\times $ (identity) have Killing vector fields 
that are everywhere null.  

To understand the Hopf fibration of the 3-sphere, it is helpful to 
begin with the observation that every Hopf circle lies on one of 
a space filling set of tori---indeed on a set of intrinsically flat tori 
of varying size and shape \cite{Roger}. We will now study our congruences 
in the same spirit, beginning with the timelike geodetic congruence 
generated by $\tilde{J}_0$ and coinciding with the $\tau$ coordinate lines. 
All the geodesics in the $\tilde{J}_0$ congruence are timelike and lie on 
a set of intrinsically flat Lorentzian tori, defined by 

\begin{equation} X^2 + Y^2 = \mbox{constant} \ . \end{equation}

\begin{figure}
       \centerline{ \hbox{
                \epsfig{figure=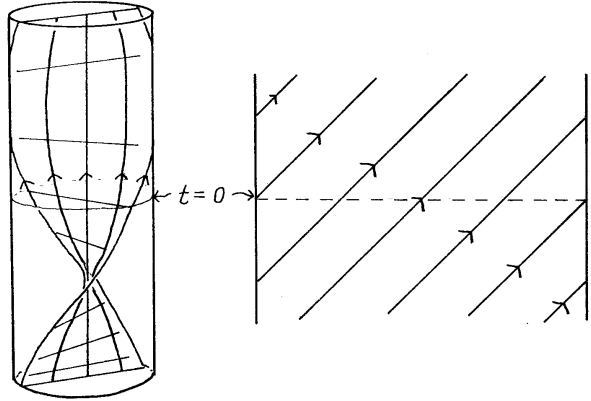,width=8cm}}}
        \caption{This picture is drawn using the sausage 
coordinates from the Appendix. It shows anti-de Sitter space as a cylinder 
(with a conformal boundary). The timelike congruence consists of timelike spirals 
ruling a set of helicoids. To the right we show that the flow becomes 
null on the boundary.}
        \label{fig:1}
\end{figure}

\noindent These are tori because (or if) anti-de Sitter space is periodic 
in the time direction. Surfaces of constant $\sigma$ are also ruled by these 
geodesics. In particular the surface 

\begin{equation} XV - VY = 0 \hspace{5mm} \Leftrightarrow \hspace{5mm} 
\sigma = 0 \end{equation}

\noindent is flat and minimal. We draw it in Fig. 1, using the sausage 
coordinates from the Appendix. In sausage coordinates the Killing vector becomes 
$\tilde{J}_0 = \partial_t + \partial_\phi $, the congruence consists a 
set of helices, and the surface $\sigma = 0$ is known as the helicoid. 
(It is a minimal surface in coordinate space too.) Note that the geodesics 
become null on the conformal boundary \scri . There are no fixed points 
anywhere.     

The spacelike congruence, generated by $J_2$ and coinciding with the 
$\sigma$ coordinate lines, is harder to draw. The first observation is that 
also these geodesics become null on \scri , although in this case there 
are two lines of fixed points; there are sources at $t - \phi = \pi/2$ 
and sinks at $t - \phi = - \pi/2$. Inside anti-de Sitter space the 
congruence is everywhere spacelike, and every Poincar\'e disk defined by 
$t = $ constant contains one member of the congruence. Surfaces 
of constant $\tau$, which are flat and minimal, 
are ruled by these geodesics but are rather hard to draw. Another surface 
that is ruled by these geodesics is the totally geodesic null surface

\begin{figure}
        \centerline{ \hbox{
                \epsfig{figure=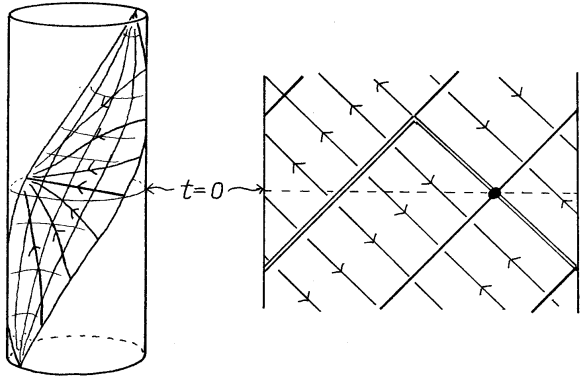,width=8cm}}}
        \caption{Again using sausage coordinates we show the null surface 
$X = V$, and how it is ruled by the spacelike congruence. To the right  
we show that the flow becomes null on the boundary, and where it has fixed points. 
There is a special point acting as a sink for all those members of the 
congruence that belong to the null surface shown.}
        \label{fig:2}
\end{figure}

\begin{equation} X = V \ . \end{equation}

\noindent In fact this surface contains every $J_2$ geodesic that goes 
to a particular sink on \scri , and is fairly easy to draw (Fig. 2). 
The one parameter family of null surfaces 

\begin{equation} X\cos{\beta} + Y\sin{\beta} + U\sin{\beta} - 
V\cos{\beta} = 0 \label{2.20} \end{equation}

\noindent provides a foliation of anti-de Sitter space with null surfaces 
ruled by the $J_2$ geodesics. 
Moreover this very family of null surfaces will be of importance 
for our discussion of squashed anti-de Sitter space later on. 

We can already see that trouble is brewing on \scri \ once we decide 
to squash our spacetime along these fibres (keeping all distances 
orthogonal to the fibres constant). On the boundary the fibres are 
changing character, from timelike/spacelike to null. Therefore 
squashing \scri \ is a different matter than squashing the interior. 
Moreover, we expect---and this is true---that the result of squashing 
all the way down to zero distance along the spacelike fibres will 
result in a two dimensional anti-de Sitter space. But this spacetime 
has a \scri \ consisting of two disconnected components. Somewhere 
along the way, something has to break. 

\vspace{1cm}

{\bf 3. Squashing, stretching, and symmetries}

\vspace{5mm}

\noindent It is time to introduce the spacetime analogues of the Berger 
sphere. We obtain them by squashing ${\bf adS}_3$ along one of the 
two congruences described in the previous section. The resulting spacetimes 
will be homogeneous but anisotropic, and we will study their symmetries 
in some detail.  

Let us consider the spacelike case first; it has some special features 
that, in the end, make this case the easiest to understand---especially if 
Fig. 2 is kept firmly in mind. The metric on squashed ${\bf adS}_3$ is 

\begin{equation} ds^2_\lambda = \frac{1}{4}\left( - \cosh^2{\omega}d\tau^2 + 
d\omega^2 + \lambda^2(d\sigma + \sinh{\omega}d\tau )^2 \right) \ , 
\label{3.1} \end{equation}

\noindent where $\lambda$ is a real squashing parameter. If we set it to 
zero we obtain the metric on ${\bf adS}_2$ in a well known coordinate 
system---which is the familiar fact that ${\bf adS}_3/{\bf R} = 
{\bf adS}_2$, analogous to the fact that ${\bf S}^3/{\bf S}^1 = 
{\bf S}^2$. Note however that---unlike its analogue for the 
3-sphere---this particular result does not 
have any straightforward higher dimensional analogue. 

Because of the squashing the isometry group is now four dimensional. 
The Lie algebra changes from $SO(2,1)\times SO(2,1)$ for the unsquashed 
spacetime to $R \times SO(2,1)$ for the squashed one; the left factor 
here gives transformations belonging to a hyperbolic conjugacy class of 
$SO(2,1)$. The question we ask is whether any Killing horizons survive. 
The answer is yes. There will be bifurcate Killing horizons coming from 
transformations of the type (hyperbolic) $\times $ (hyperbolic), although 
they will be less numerous than they were in anti-de Sitter space. The 
degenerate Killing horizons that were present in the unsquashed case 
are no longer with us, since they came from transformations of the 
type (parabolic) $\times $ (parabolic). But there were also totally null 
Killing vector fields in anti-de Sitter space, coming from transformations 
of the type (identity) $\times $ (parabolic). Once we have done the 
squashing this will give us a supply of degenerate Killing horizons, 
as a replacement for those that were lost. 

But we do not have to rely on any previous results here. A short calculation 
verifies that the most general Killing vector field that has a spacelike 
curve of fixed points is (up to scale) 

\begin{equation} \xi = J_2 + a\tilde{J}_1 + b\tilde{J}_2 + c\tilde{J}_0 
\ , \end{equation}

\noindent where the real numbers $a$, $b$, $c$ obey 

\begin{equation} a^2 + b^2 - c^2 = 1 \ . \end{equation}   

\noindent This is a timelike surface in the group manifold of $SO(2,1)$. 
The fixed points occur at 

\begin{equation} \sinh{\omega} = c \ , \hspace{6mm} \cos{\tau} = 
\frac{b}{\sqrt{a^2 + b^2}} \hspace{4mm} \sin{\tau} = - \frac{a}
{\sqrt{a^2 + b^2}} \ . \end{equation}

\noindent These curves are precisely the fibres along which we are 
squashing. They are also bifurcation curves for bifurcate 
Killing horizons. Hence squashed anti-de Sitter space contains a two 
parameter family of bifurcate Killing horizons. The unsquashed 
spacetime has more: in anti-de Sitter space itself every spacelike 
geodesic is a bifurcation line for some Killing horizon.

If we pick an example in this class, we find that 

\begin{equation} ||J_2 + \tilde{J}_2||^2 = (1 - \cos{\tau}\cosh{\omega})
\left( \lambda^2 + 1 - (\lambda^2 - 1)\cos{\tau}\cosh{\omega}\right) 
\ . \end{equation}

\noindent The surface gravity $\kappa$ is given by $|\kappa | = 2$. A 
feature that arises only in the squashed case is that there are actually 
two surfaces where the norm vanishes, but only one of them is a Killing 
horizon---the other is a timelike surface. 

A one parameter family of degenerate Killing horizons arise from the 
Killing vectors 
 
\begin{equation} \tilde{\xi} (\alpha) = 
\cos{\alpha}\tilde{J}_1 + \sin{\alpha}\tilde{J}_2 
+ \tilde{J}_0 \ . \end{equation}

\noindent This time the norm is

\begin{equation} ||\tilde{\xi} (\alpha)||^2 = (\lambda^2 - 1)(\sinh{\omega} + 
\sin{(\tau - \alpha)}\cosh{\omega})^2 \ . \label{28} \end{equation}

\noindent In ${\bf adS}_3$ these Killing vectors are everywhere null. 
In the squashed case ($\lambda^2 < 1$) they are timelike except for a 
degenerate Killing horizon where the norm vanishes, and in the stretched 
case ($\lambda^2 > 1$) they are spacelike again except for a degenerate 
Killing horizon. In anti-de 
Sitter space itself this family of null surfaces is identical to the 
family given in eq. (\ref{2.20}), if we set $\alpha = 2\beta + \pi$. 
In the anti-de Sitter case there are additional degenerate Killing horizons 
that disappear when we squash or stretch.

Since we are primarily interested in Killing horizons because they are 
totally geodesic null surfaces, it is enough to consider degenerate 
horizons---the bifurcate ones do not contribute anything new in 
this way.

Next we squash or stretch along the timelike congruence. Then the 
metric is 

\begin{equation} ds_{\lambda}^2 = \frac{1}{4}\left( - \lambda^2
(d\tau - \sinh{\omega}d\sigma )^2 + d\omega^2 + 
\cosh^2{\omega}d\sigma^2 \right) \ . \label{3.29} \end{equation}

\noindent Setting $\lambda = 0$ results in the metric on the hyperbolic plane, 
expressing the well known fact that ${\bf adS}_3/{\bf S}^1 = {\bf H}^2$. 
In general the symmetry group is $SO(2)\times SO(2,1)$. This means that 
there are no bifurcate Killing horizons anymore. There are no degenerate 
horizons either. We may try 

\begin{equation} \xi = J_2 + J_0 \ , \end{equation}

\noindent which is everywhere null in anti-de Sitter space. After (timelike) 
squashing we obtain

\begin{equation} ||\xi ||^2 = - (\lambda^2 - 1)(\cosh{\omega}
\cosh{\sigma} - \sinh{\omega})^2 \ . \end{equation}

\noindent This is timelike or spacelike, depending on whether we squash or 
stretch. 

\vspace{1cm}

{\bf 4. Null geodesics}

\vspace{5mm}

\noindent Our spacetimes have enough symmetries to ensure that the geodesic 
equation can be separated. It is particularly interesting to take a look 
at the equations for null geodesics, because there is a surprise waiting. 
We will discuss the case of spacelike squashing in some detail, and comment 
briefly on timelike squashing at the end. For a related discussion, 
including some interesting observations on timelike geodesics, see 
Bardeen and Horowitz \cite{Gary}. 

We begin by introducing the convenient coordinate

\begin{equation} {\rm w} = \sinh{\omega} \ . \end{equation}

\noindent Using it, it is straightforward to bring the equations for a 
null geodesic with respect to the metric (\ref{3.1}) to the form 

\begin{equation} \dot{\tau} = \frac{1 + \lambda {\rm w}\sin{\phi}}
{1 + {\rm w}^2} \end{equation}

\begin{equation} \dot{\sigma} = \frac{\sin{\phi}}{\lambda} - 
\frac{\rm w}{1+{\rm w}^2}(1+\lambda {\rm w}\sin{\phi}) \end{equation}

\begin{equation} \dot{\rm w}^2 = \cos^2{\phi} + 2\lambda {\rm w}\sin{\phi}
+ (\lambda^2 - 1){\rm w}^2\sin^2{\phi} \ . \end{equation}

\noindent Without loss of generality we have chosen the integration constants 
to ensure that the geodesic passes the origin of our coordinate system. The 
first observation is that if $\sin{\phi} = 0$ then $\lambda$ drops out of 
the equations; this corresponds to a null geodesic that is everywhere 
orthogonal to the squashing direction. Such geodesics are unaffected 
by any squashing (or stretching). This is actually true also for spacelike 
and timelike geodesics orthogonal to the squashing direction, as well as 
for the spacelike geodesics that are parallel to this direction. Note 
also that we have come across null geodesics orthogonal to the squashing 
direction once before---they rule the Killing horizons depicted in Fig. 2.   

Let us now assume that $\sin{\phi} \neq 0$. Asymptotically, that is 
to say for large values of ${\rm w}$, we obtain

\begin{equation} \dot{\tau} \sim \frac{\lambda \sin{\phi}}{\rm w} \end{equation}

\begin{equation} \dot{\sigma} \sim (1-\lambda^2)\frac{\sin{\phi}}{\lambda} 
\end{equation}

\begin{equation} \dot{\rm w}^2 \sim (\lambda^2 - 1){\rm w}^2\sin^2{\phi} \ . 
\end{equation}

\noindent Evidently it is possible to reach arbitrarily large values of ${\rm w}$ 
only if $\lambda^2 \geq 1$, that is to say only for stretching, not for 
squashing. This is the surprise that we were referring to. 

The explicit solution for ${\rm w}(s)$ can be written down, but is not very 
illuminating---we get the expected oscillatory behaviour for squashing, while 
stretching gives an exponentially growing function. To go on, when 
$\lambda^2 > 1$ we see that 

\begin{equation} {\rm w} \rightarrow \infty \hspace{5mm} \Rightarrow 
\hspace{5mm} \sigma \rightarrow \pm \infty \ . \end{equation}

\noindent This is a very different kind of behaviour from that occurring 
in the unstretched anti-de Sitter case. In effect, asymptotically the null 
geodesics are lining up with the null geodesics that rule the Killing horizons 
described in the previous section. The implications of this will be discussed 
in section 6. 

For the case of timelike squashing fibres detailed results are available in the 
literature already. This is because, by adding an extra flat direction, 
and specializing the stretching parameter to $\lambda^2 = 2$ \cite{Spindel}, 
the resulting 3+1 dimensional spacetime is the famous G\"odel solution. 
An elegant review of its null geodesics has been given by Ostv\'ath and 
Schucking \cite{Schucking}; to follow them we use the coordinate system 
given just before eq. (\ref{90}) in the Appendix, and perform the further 
coordinate changes 

\begin{equation} T = \frac{1}{2}(\psi - \phi ) \ , \hspace{6mm} 
\cosh{\theta} = \frac{1+R^2}{1-R^2} \ , \hspace{3mm} \sinh{\theta} 
= \frac{2R}{1-R^2} \ , \hspace{3mm} 0 < R < 1 \ . \end{equation}

\noindent This brings the metric to the form 

\begin{equation} ds^2 = \frac{1}{(1-R^2)^2}\left[ - \lambda^2\left( 
(1-R^2)dT - R^2d\phi\right)^2 + dR^2 + R^2d\phi^2 \right] \ . 
\end{equation}
  
\noindent When we study null geodesic paths we may ignore the conformal 
factor in front of the metric. What one finds \cite{Schucking} is that 
a null geodesic through the origin of our coordinate system obeys 

\begin{equation} R = \frac{1}{\lambda}\sin{(\phi + \phi_0)} \ . \end{equation}

\noindent Setting $\phi_0 = 0$ and trading $R$ and $\phi$ for Cartesian 
coordinates on the Poincar\'e disk gives 

\begin{equation} \sqrt{x^2 + y^2} = \frac{1}{\lambda}\frac{y}{\sqrt{x^2 + y^2}} 
\hspace{5mm} \Leftrightarrow \hspace{5mm} x^2 + \left( y - \frac{1}{2\lambda}
\right)^2 = \left( \frac{1}{2\lambda}\right)^2 \ . \end{equation}

\noindent This is a circle. The family of null geodesics through the origin, 
projected down to the Poincar\'e disk, are circles whose envelope is a circle 
with radius $1/\lambda$. 

Thus when $\lambda^2 > 1$ all null geodesics are confined to the interior of 
stretched anti-de Sitter space. When $R = 1$ the conformal factor in front of 
the metric diverges, so that what happens when $\lambda = 1$ is that the null 
geodesics touch \scri \ but they slow down there in such a way that there 
is no turning back. In the squashed case ($\lambda^2 < 1$) the null geodesics 
escape. Detailed 
examination shows that the squashed case differs from the anti-de Sitter 
case in that the null geodesics reach arbitrarily large values of the 
coordinate $T$, rather than end up at a finite value of $T$ as they do in 
anti-de Sitter space. 
 
We may further observe that 

\begin{equation} || \partial_\phi||^2 = \frac{R^2}{(1-R^2)^2}(1- \lambda^2R^2) 
\ . \end{equation}

\noindent Hence $\lambda^2 > 1$ implies that there are closed timelike curves 
beyond the envelope of the null geodesics; this, of course, was one of G\"odel's 
main points. In the squashed case no such thing happens. Indeed 

\begin{equation} dT^2 = 0 \hspace{5mm} \Rightarrow \hspace{5mm} 
ds^2 = \frac{1}{(1-R^2)^2}\left( dR^2 + R^2(1-\lambda^2R^2)d\phi^2\right) 
\ . \end{equation}

\noindent Thus when $\lambda^2 < 1$ the coordinate $T$ serves as a global 
time function, and there can be no CTCs, unless this direction is compactified. 
There are no CTCs in the case of spacelike squashing or stretching either.

In the next section we will observe another key difference between 
squashing and stretching, and between spacelike and timelike fibres.

\vspace{1cm}

{\bf 5. The Einstein--Weyl equations}

\vspace{5mm}

\noindent The Einstein--Weyl equations are a conformally invariant set of 
equations involving a metric tensor and a vector potential. They were 
introduced by Weyl in an attempt to unify gravitation and electromagnetism 
\cite{Weyl}; his theory failed but left a valuable legacy. We will give the basic 
facts only; for a more complete summary see Pedersen and Tod \cite{Pedersen}. 

By definition, a Weyl space is a manifold equipped by a metric 
$g_{ab}$, a one-form $\omega_a$, and a connection---known as the Weyl 
connection---which are compatible in the sense that 

\begin{equation} D_ag_{bc} = \omega_ag_{bc} \ . \end{equation}

\noindent The solution is  

\begin{equation} D_aV^b = \nabla_aV^b + {\gamma}^b_{\ ac}V^c 
\label{47} \end{equation}

\noindent where $\nabla_a$ is defined by the usual metric compatible 
Levi-Civita connection and   

\begin{equation} {\gamma}^b_{\ ac} = - \frac{1}{2}(\delta^b_a\omega_c + 
\delta^b_c\omega_a - g_{ac}g^{bd}\omega_d) \ . \label{48} \end{equation}

\noindent Given a Weyl space, the pair $(g^{\prime}_{ab}, \omega^{\prime}_c) 
= (e^{\Omega}g_{ab}, \omega_c + \partial_c{\Omega})$ defines a Weyl space too. 

The Weyl connection has a curvature tensor defined by 

\begin{equation} [D_a, D_b]V_c = W_{abc}^{\ \ \ d}V_d \ . \end{equation}

\noindent We also define 

\begin{equation} W_{ab} = W_{acb}^{\ \ \ c} \hspace{5mm} \mbox{and} 
\hspace{5mm} W = g^{ab}W_{ab} \ . \end{equation}

\noindent Note that $W_{ab}$ is not symmetric in general. A calculation 
shows that 

\begin{eqnarray} W_{(ab)} & = & R_{ab} + \frac{1}{2}\nabla_{(a}\omega_{b)} + 
\frac{1}{4}\omega_a\omega_b + g_{ab}\left( \frac{1}{2}\nabla_c\omega^c - 
\frac{1}{4}\omega^2\right) \\
\ \nonumber \\
W_{[ab]} & = & \frac{3}{2} \nabla_{[a}\omega_{b]} \equiv \frac{3}{2}F_{ab} 
\ , \end{eqnarray}

\noindent where square and round brackets denote anti-symmetrization and 
symmetrization, respectively. Notice the definition of $F_{ab}$. 
It is easy to see that 

\begin{equation} W_{ab}^{\ \ cd} = W_{ab}^{\ \ [cd]} + F_{ab}g^{cd} \ , 
\end{equation}

\noindent and moreover---because we are in 3 dimensions---

\begin{equation} W_{ab}^{\ \ cd} = \epsilon_{abe}\epsilon^{cdf}\left( 
\frac{1}{2}\delta_f^eW - W_f^{\ e} - F_f^{\ e}\right) \ . \end{equation}

\noindent By definition a three dimensional Einstein--Weyl space obeys

\begin{equation} W_{(ab)} = \frac{1}{3}Wg_{ab} \ . \end{equation}  

\noindent For the ordinary Ricci tensor this implies that 

\begin{equation} R_{ab} + \frac{1}{2}\nabla_{(a}\omega_{b)} + 
\frac{1}{4}\omega_a\omega_b = g_{ab}\left( \frac{1}{3}W - \frac{1}{2}
\nabla_c\omega^c + \frac{1}{4}\omega^2\right) = \Lambda g_{ab} \ , 
\label{37} \end{equation}

\noindent where $\Lambda$ is some function. In the Einstein case the 
Bianchi identities force $\Lambda$ to equal a constant, but this is 
no longer true here. Unlike the Einstein equations, the Einstein--Weyl 
equations admit an infinitude of locally inequivalent solutions in 3 
dimensions. It was shown by Cartan that these solutions can be specified 
by four arbitrary functions of two variables \cite{Cartan}. 
 
Does squashed anti-de Sitter space obey the Einstein--Weyl equations? 
For the spacelike case, we start with the metric (\ref{3.1}) and compute 
the Ricci tensor. We must then find a one-form $\omega_a$ such that 
eq. (\ref{37}) holds for some $\Lambda $. What we actually find is that 

\begin{equation} R_{ab} + \lambda^2(\lambda^2 - 1)\xi_a \xi_b = 
2(\lambda^2 - 2) g_{ab} \ , \label{4.12} \end{equation}

\noindent where the one-form $\xi^a$ is the Killing vector field that 
defines the squashing, 

\begin{equation} \xi_a = \nabla_a\sigma + \sinh{\omega}\nabla_a\tau \ 
\hspace{5mm} \Rightarrow \hspace{5mm} \nabla_{(a}\xi_{b)} = 0 
\ . \end{equation}

\noindent Therefore we obtain a solution of the Einstein--Weyl equation 
if we perform the rescaling

\begin{equation} \frac{1}{2}\omega_a \equiv \sqrt{\lambda^2(\lambda^2 - 1)}
\xi_a \ . \end{equation}

\noindent Curiously a real solution is obtained only for $\lambda^2 \geq 1$, 
that is to say if we stretch anti-de Sitter space, but not if we squash it.   
For timelike squashing, we obtain a real solution when we squash but not 
when we stretch; this is also true for the Riemannian Berger sphere 
\cite{Pedersen}. We do not fully understand why this should be so. We 
observe that, in anti-de Sitter space, spacelike geodesics tend to diverge, 
and timelike geodesics tend to converge. Geodesics on the 3-sphere tend to 
converge as well. Perhaps more to the point, in the previous section 
we noted that null geodesics behave very differently depending on whether 
the spacetime is squashed or streched. 

With the Weyl connection in hand we can define a new notion of geodesic 
curves. We will continue to refer to geodesics with respect to our chosen 
metrics as ``geodesics'', while geodesics with respect to the Weyl connection 
will be called ``Weyl geodesics''. Cartan proved that---at least after 
complexification---a three dimensional Einstein--Weyl space admits a two 
parameter family of null hypersurfaces that are totally geodesic with respect 
to the Weyl connection. It is this two dimensional space that is used as a 
mini-twistor space by Jones and Tod \cite{Jones}. In the anti-de Sitter case 
the mini-twistor space can be identified with \scri , the conformal boundary 
of spacetime (since then the two notions of ``geodesic'' coincide, and any 
point on \scri \ can be regarded as the vertex of a past light cone which 
is totally geodesic---this is true for the de Sitter case as well). It would 
be interesting to see explicitly what these null surfaces are in the 
squashed cases. We do not know, but we will show that the degenerate Killing 
horizons that we found for spacelike squashing, eq. (\ref{28}), do belong 
to this set. 

Following Pedersen and Tod \cite{Pedersen}, let us analyze the Weyl geodesics. 
From eqs. (\ref{47}--\ref{48} ) it is seen that they obey 

\begin{equation} \dot{x}^bD_b\dot{x}^a = \dot{x}^b\nabla_b\dot{x}^a - 
E\dot{x}^a + \frac{1}{2}\dot{x}^2\omega^a = \alpha \dot{x}^a \ , \end{equation}

\noindent where 

\begin{equation} E \equiv \dot{x}^a\omega_a \end{equation}

\noindent and $\alpha$ depends on how $x^a(s)$ is parametrized. The 
first observation is that null geodesics are null geodesics with respect 
to the Weyl connection too; only the parametrization differs. To study 
the remaining cases it is convenient to parametrize the curves using 
arc length ($\dot{x}^2 = \pm 1$). It turns out that this requires that 
$\alpha = -E/2$, and then we find that the Weyl geodesics obey

\begin{equation} \dot{x}^b\nabla_b\dot{x}^a + \frac{1}{2}\dot{x}^2
\omega^a = \frac{1}{2}E\dot{x}^a \ . \end{equation}

\noindent There is a ``force'' directed along $\omega^a$. 

In the cases that we are interested in $\omega^a$ is a Killing vector 
of constant norm (and hence the tangent vector of a geodesic). This 
simplifies matters considerably, and leads to the equation

\begin{equation} 2\dot{E} = E^2 - \omega^2\dot{x}^2 \ . \end{equation}

\noindent This equation can be solved. For spacelike stretching, or 
more generally when $\omega^a$ is spacelike, we find that spacelike 
Weyl geodesics obey $E^2 \rightarrow \omega^2$. This then implies 
that \cite{Pedersen} 

\begin{equation} \dot{x}^a \rightarrow \frac{\omega^a}{\sqrt{\omega^2}} 
\ . \end{equation}

\noindent Similarly we find for timelike squashing that timelike geodesics 
obey $\dot{x}^a \rightarrow \omega^a/\sqrt{-\omega^2}$. Asymptotically, 
these Weyl geodesics line up with the squashing direction. 
  
It is by now evident that the degenerate Killing horizons that we found for 
spacelike squashing are totally geodesic with respect to the Weyl 
connection. Being Killing horizons they are totally geodesic with 
respect to the metric connection. Null geodesics coincide for both 
connections, and the spacelike Weyl geodesic deviate from the spacelike 
metric geodesics in the direction of the squashing field---which as 
we know is tangential to the Killing horizons. (This argument depends 
critically on the fact that the squashing field is tangential to the 
Killing horizon. We would not be surprised to learn that these are the 
only null surfaces in our spacetimes that are totally geodesic in the 
ordinary sense.)

For any Einstein--Weyl space with a spacelike $\omega_a$ we observe that 
every null geodesic belongs to some null surface that is totally 
geodesic with respect to the Weyl connection \cite{Pedersen}. Since all 
spacelike Weyl geodesics eventually line up with $\omega_a$, this has 
consequences for the behaviour of the null geodesics ``close to 
infinity''---a phrase that we will examine in more detail in the next 
section. 

\vspace{1cm}

{\bf 6. Conformal compactification?}

\vspace{5mm}

\noindent We will now point our moral. It concerns the fragility of \scri , 
the conformal boundary of an Einstein space. Recall that the idea---in 
outline!---is to perform a conformal transformation 

\begin{equation} g_{ab} \rightarrow \hat{g}_{ab} = \Omega^2g_{ab} \ , 
\end{equation} 

\noindent and to choose the conformal factor $\Omega $ in such a way that 
the original manifold can be regarded as sitting inside a hypersurface in 
an extended conformally related spacetime. This hypersurface is  
defined by $\Omega = 0$, and is called \scri . The affine parameters 
on null geodesics will be finite when they reach \scri , if defined using 
$\hat{g}_{ab}$, although they diverge when defined using the original metric. 
For Einstein spaces it is known that, whenever it exists, \scri \ is a null 
hypersurface if the cosmological constant vanishes, while it is timelike 
(spacelike) for negative (positive) cosmological constant. But the argument 
that leads to this conclusion \cite{Penrose} relies on the Einstein 
equations, and becomes void for the cases we study. 

For our purposes we will insist that \scri \ is a surface with 
almost everywhere defined normal vector in a conformally related spacetime, 
and that every point on \scri \ can be regarded as the vertex of a 
past directed light cone, with a non-zero fraction of its generators 
belonging to the original spacetime. 
 
Let us first recall the conformal compactification of ordinary anti-de Sitter 
space, using our unusual coordinates. A standard choice of conformal factor 
is \cite{Holst}

\begin{equation} \Omega^2 = 1/(U^2 + V^2) = 2/(\cosh{\omega}
\cosh{\sigma} + 1) \ . \end{equation}

\noindent Using it, the conformally related metric becomes

\begin{equation} d\hat{s}^2 = \Omega^2ds^2 = \frac{1}
{2(\cosh{\omega}\cosh{\sigma} + 1)}\left( - d\tau^2 + d\omega^2 + d\sigma^2 
+ 2\sinh{\omega}d\tau d\sigma \right) \ . \end{equation}

\noindent We now take the limit $\omega \rightarrow \infty$ to obtain the 
metric on \scri . Actually this will give us ``one half'' of 
\scri \ only ; the other half sits at $\omega \rightarrow - \infty $. 
The metric on \scri \ is found to be 

\begin{equation} d\hat{s}^2 = \frac{d\tau d\sigma}{\cosh{\sigma}} \ . 
\end{equation}

\noindent Therefore $\tau$ and $\sigma$ are null coordinates on \scri . 
A more convenient choice of coordinates are $u$ and $v$, where  

\begin{equation} \tan{u} = - \sinh{\sigma} \hspace{8mm} v = \tau \ . 
\end{equation}

\noindent We see that $\sigma = \pm \infty $ is a null line on \scri , 
dividing it into two halves. The metric on \scri \ is 

\begin{equation} d\hat{s}^2 = - du dv \ . \end{equation}

\noindent The spacetime Killing vectors, restricted to \scri , generate 
conformal isometries of this flat Lorentzian metric. One can show that 
this \scri \ is a timelike surface in a 2+1 dimensional Einstein universe.   

Squashed or stretched anti-de Sitter space cannot work quite like this. 
This actually follows from the discussion in section 4. For spacelike 
stretching the (ideal) end points of the set of null geodesics form 
a one dimensional set, and therefore they cannot form a \scri . For 
spacelike squashing the case is less clear, but we suggest it may 
be even worse. For timelike stretching the null geodesics are trapped 
inside spacetime; although strictly speaking this does not exclude the 
existence of a spacelike \scri \ at future infinity, we expect the 
end points to form a zero dimensional set. 
For timelike squashing the situation is again less clear, but it seems 
likely that this case is similar to that of spacelike stretching. 
So we conclude that there can be no \scri \ in any case. 

Let us now proceed in a direct manner to see if we arrive at the same 
conclusion. A look at the metric in eq. (\ref{3.1}) shows that, 
as soon as $\lambda \neq 1$, the asymptotic dependence on $\omega$ changes 
dramatically. To get a finite expression we must choose something like 

\begin{equation} \Omega^2 = 4/\sinh^2{\omega} \end{equation}

\noindent (up to some factor that remains finite in the limit). Then 
 
\begin{equation} d\hat{s}^2_\lambda = \Omega^2ds^2_\lambda = 
\frac{1}{\sinh^2{\omega}}\left( - \cosh^2{\omega}d\tau^2 + 
d\omega^2 + \lambda^2(d\sigma + \sinh{\omega}d\tau )^2 \right) 
\end{equation}

\noindent The hypersurface $\Omega = 0$ is supposed to sit at $\omega 
\rightarrow \infty $. In the unsquashed case this was a Lorentzian cylinder. 
But in the squashed case we obtain

\begin{equation} \lim_{\omega \rightarrow \pm \infty} d\hat{s}^2_\lambda = 
- (1 - \lambda^2)d\tau^2 + 0\cdot d\omega^2 + 
0 \cdot d\sigma^2 \ . \end{equation}

\noindent This is a degenerate metric, so it would seem as if the squashing 
has caused the conformal boundary to become null. 
 
But actually it is worse than this. If $\hat{R}$ denotes the Ricci scalar 
of the conformally related metric $\hat{g}_{ab}$ it will be true that 

\begin{equation} \hat{R} = \frac{1}{\Omega^2}\left( R - 4g^{ab}\nabla_a
\nabla_b\ln{\Omega} - 2g^{ab}\nabla_a\ln{\Omega}\nabla_b\ln{\Omega} \right) 
\ . \end{equation}

\noindent But we know from eq. (\ref{4.12}) that 

\begin{equation} R = 2(\lambda^2 - 4) \ . \end{equation}

\noindent It is then clear that $\hat{R}$ will diverge when $\omega 
\rightarrow \infty$, unless the asymptotic behaviour of $\Omega$ is 
carefully adjusted. The choice $\Omega \sim \exp{(-\omega/2)}$ that we 
made for anti-de Sitter space leads to a finite $\hat{R}$, but for 
all $\lambda > 0$ the choice $\Omega \sim \exp{(-\omega)}$ gives a 
curvature singularity instead of a well defined conformal boundary 
at infinity.  

This argument has its flaws. Although $ds^2$ is a scalar, it does not really 
have an invariant meaning. It is simply the length squared of a vector that 
in a particular coordinate system has the finite components $d\tau$, 
$d\omega$, $d\sigma$. However, since our understanding of the null geodesics 
led us to the same conclusion, we dare to claim that there simply cannot be 
any conformal compactification of squashed or stretched anti-de Sitter 
space, in any conventional sense. In itself this is not a very surprising 
conclusion since there is more to conformal compactification than just 
Lorentzian geometry. 

The moral is that if we deviate from Einstein's equations, we court 
disaster. In the Einstein--Weyl cases one might think that the existence 
of a two parameter family of totally Weyl geodesic null surfaces should 
somehow guarantee the existence of \scri , but this is not so. For the 
case of spacelike squashing the problem is that---as shown in section 
5---the spacelike geodesics that these null surfaces contain will tend 
asymptotically to go in the squashing direction. In a sense then there 
are not enough distinct such surfaces ``at infinity''.    

\vspace{1cm}

{\bf 7. The extremal Kerr black hole}

\vspace{5mm}

\noindent Now for a more directly physical application. In Boyer--Lindquist 
coordinates the Kerr solution, the unique solution describing a spinning 
black hole in 3+1 dimensions, takes the form 

\begin{equation} ds^2 = - e^{2\nu}d\tilde{t}^2 + \frac{\rho^2d\tilde{r}^2}
{\Delta} + \rho^2d\theta^2 + \frac{\Delta \sin^2{\theta}}{e^{2\nu}}
(d\tilde{\phi} - \Omega d\tilde{t})^2 \end{equation} 

\noindent where 

\begin{equation} \rho^2 \equiv \tilde{r}^2 + a^2\cos^2{\theta} \ , \hspace{7mm} 
\Delta \equiv \tilde{r}^2 - 2M\tilde{r} + a^2 \ , \end{equation}

\begin{equation} e^{2\nu} \equiv \frac{\Delta \rho^2}{(\tilde{r}^2 + a^2)^2 
- \Delta a^2\sin^2{\theta}} \ , \hspace{4mm} \Omega \equiv 
\frac{2M\tilde{r}ae^{2\nu}}{\Delta \rho^2} \ . \end{equation}

\noindent The mass of this black hole equals $M$, and its angular momentum 
$J = aM$. From now on we will be interested in the extremal limit $J = M^2$. 
Then the horizon is at $\tilde{r} = M$, and the angular velocity of 
the horizon is $\Omega = 1/2M$. 

At constant $\tilde{t}$ the spatial distance to the extremal horizon 
is infinite. This is reminiscent of the extremal Reissner--Nordstr\"om 
black hole. In the latter case it is well known that the near horizon 
geometry of the extremal black hole is ${\bf adS}_2\times {\bf S}^2$, 
and the event horizon sits at a degenerate Killing horizon in ${\bf adS}_2$. 
Bardeen and Horowitz \cite{Gary} have pointed out that the near horizon 
limit of the extremal Kerr black hole is quite simple too. To arrive 
at it, we set 

\begin{equation} \tilde{r} = M + kr \ , \hspace{4mm} \tilde{t} = 
2M^2t/k \ , \hspace{5mm} \tilde{\phi} = \phi + Mt/k \ . \end{equation}

\noindent The Killing vector $\partial_t$ rules the horizon. 
Next we take the limit $k \rightarrow 0$. Bardeen and Horowitz 
follow this up by a coordinate transformation that allows them to 
analytically continue the spacetime, so that it becomes geodesically 
complete. More precisely

\begin{equation} r = \cosh{\omega}\cos{\tau} + \sinh{\omega} \ , 
\hspace{4mm} t = \frac{\sinh{\omega}\sin{\tau}}{r} \ , 
\hspace{4mm} \phi = \sigma + f(\tau, \omega) \ , \end{equation}

\noindent where the function $f$ is chosen so that the metric simplifies. 
The final result \cite{Gary} is a spacetime with the metric

\begin{eqnarray} ds^2 = M^2(1+\cos^2{\theta}) 
\left( - \cosh^2{\omega}d\tau^2 + d\omega^2 + d\theta^2 \right) 
+ \nonumber \\
\ \label{Gary} \\
\hspace{25mm} + \frac{4M^2\sin^2{\theta}}{1+\cos^2{\theta}}
(d\sigma + \sinh{\omega}d\tau )^2 \ . \nonumber \end{eqnarray}

\noindent The coordinates $\tau$ and $\omega$ run from $- \infty$ to 
$+ \infty $, while $\sigma$ is a periodic coordinate. 
This spacetime is in itself a vacuum solution of Einstein's 
equations \cite{Meinel}.

The reason why we bring this up is evidently that the near 
horizon geometry at fixed $\theta$ is precisely a 2+1 dimensional 
anti-de Sitter space, squashed or stretched in a $\theta$ dependent 
way. See eq. (\ref{3.1}). The 
only new thing here is that periodicity in $\sigma$ has been 
imposed, changing the symmetry group to $U(1)\times SO(2,1)$. 

At $\theta = 0$ the metric (\ref{Gary}) is just the metric on 
${\bf adS}_2$. When 

\begin{equation} \cos^2{\theta} = 2\sqrt{3} - 3 \ , \end{equation}

\noindent that is roughly at $\theta = 47.1^{\circ}$, the metric 
(\ref{Gary}) is ordinary 2+1 dimensional anti-de Sitter space. 
Closer to the equatorial plane we have a stretched rather than 
a squashed anti-de Sitter space. In all cases the original event 
horizon sits at 

\begin{equation} r = \cosh{\omega}\cos{\tau} + \sinh{\omega} = 0 \ . 
\label{horizon} \end{equation}

\noindent Evidently this coincides with one of the degenerate Killing 
horizons discussed in section 3. In this respect the near horizon geometry 
of the Kerr black hole resembles that of its Reissner--Nordstr\"om 
counterpart. 

\vspace{1cm}

{\bf 8. Conclusions and open questions}

\vspace{5mm}

\noindent 2+1 dimensional anti-de Sitter space can be squashed along 
fibres that are either spacelike or timelike. The symmetries and null 
geodesics of the resulting spacetimes were studied in detail. At 
several points of the discussion we even went into great detail,  
because we feel that squashed anti-de Sitter space has the 
potential for being exploited in many contexts, 
and collecting background information in one place (this paper) seemed 
like a good idea. For spacelike squashing or stretching we 
found that the spacetime contains a one parameter family of degenerate 
Killing horizons that are totally geodesic also with respect to the Weyl 
connection, that the behaviour of null geodesics is qualitatively 
different depending on whether we squash or stretch, and that the 
Einstein--Weyl condition holds only for stretching. For timelike 
squashing or stretching there are no Killing 
horizons, null geodesics are trapped inside the spacetime if we stretch 
it but not if we squash it, and the Einstein--Weyl 
condition holds only for squashing. In all cases we argued that squashing 
or stretching means that the spacetime does not admit a conformal boundary 
on which null geodesics end. 

We find the following open questions of interest: First, we did not 
state our claim about the absence of a conformal boundary as a theorem. 
Our arguments convinced us but are slightly less than watertight. 
Second, we did not explicitly construct the two parameter family 
of null surfaces totally geodesic with respect to the Weyl connection. 
For timelike squashing we did not even find any examples. Third, we 
suggest that it should be possible to formulate a 
useful theorem concerning the behaviour of null geodesics in 
any Einstein--Weyl spacetime where $\omega_a$ is a spacelike Killing 
vector field.   
   
Finally 2+1 dimensional anti-de Sitter space, squashed or stretched 
along compactified spacelike fibres, appears in the near horizon geometry 
of the extremal Kerr black hole. The constellation of Sagittarius seems 
to contain a black 
hole with $J/M^2 = 0.9939^{+0.0026}_{-0.0074}$ \cite{astr}, so we 
conclude that to a very good approximation there are squashed anti-de 
Sitter spaces at the centre of the Milky Way.
 
\vspace{1cm}

{\bf Acknowledgements}

\vspace{5mm}

\noindent We thank Johan Br\"annlund for helping to start this project, 
Paul Tod for electronic comments, S\"oren Holst and James Vickers 
for verbal advice, Jan \AA man for some checks, and St\'ephane 
Detournay for correcting a bad oversight in the first version of this 
paper.

\vspace{1cm}

{\bf Appendix}

\vspace{5mm}

\noindent Here we give details about three useful coordinate systems on 
${\bf adS}_3$. First, the sausage coordinates $(t, r, \phi)$ are defined by 

\begin{equation} X = \frac{2r}{1-r^2}\cos{\phi} \hspace{7mm} 
Y = \frac{2r}{1-r^2}\sin{\phi} \end{equation}

\begin{equation} U = \frac{1+r^2}{1-r^2}\cos{t} \hspace{8mm} 
V = \frac{1+r^2}{1-r^2}\sin{t} \ . \end{equation}

\noindent Then the metric takes the form

\begin{equation} ds^2 = - \left( \frac{1+r^2 }{1-r^2}\right)^2dt^2 + 
\frac{4}{(1-r^2)^2}(dr^2 + r^2d\phi^2) \ . \end{equation}

\noindent This is useful for visualization; it leads to a picture 
of ${\bf adS}_3$ as a salami sliced with Poincar\'e disks.  

For the calculations done in this paper the coordinates $(\tau, \omega, 
\sigma)$ are more useful \cite{Marc}. 
They were presented in section 2. Let us supplement that description with 
the explicit equations

\begin{eqnarray} 
X & = & \cos{\frac{\tau}{2}}\sinh{\frac{\omega}{2}}\cosh{\frac{\sigma}{2}} -   
\sin{\frac{\tau}{2}}\cosh{\frac{\omega}{2}}\sinh{\frac{\sigma}{2}} \\
\ \nonumber \\
 Y & = & \sin{\frac{\tau}{2}}\sinh{\frac{\omega}{2}}\cosh{\frac{\sigma}{2}} +   
\cos{\frac{\tau}{2}}\cosh{\frac{\omega}{2}}\sinh{\frac{\sigma}{2}} \\
\ \nonumber \\
U & = & \cos{\frac{\tau}{2}}\cosh{\frac{\omega}{2}}\cosh{\frac{\sigma}{2}} +   
\sin{\frac{\tau}{2}}\sinh{\frac{\omega}{2}}\sinh{\frac{\sigma}{2}} \\
\ \nonumber \\
V & = & \sin{\frac{\tau}{2}}\cosh{\frac{\omega}{2}}\cosh{\frac{\sigma}{2}} -   
\cos{\frac{\tau}{2}}\sinh{\frac{\omega}{2}}\sinh{\frac{\sigma}{2}} \ . 
\end{eqnarray}

Finally, timelike squashing was briefly discussed by Pedersen and Tod
\cite{Pedersen}. They used intrinsic coordinates related to ours by 

\begin{equation} \tanh{\sigma} = \tanh{\theta}\sin{\phi} \ , \hspace{4mm} 
\sinh{\omega} = \sinh{\theta}\cos{\phi} \ , \hspace{4mm} 
\tau = \psi + f(\theta, \phi ) \ , \label{90} \end{equation}

\noindent where $f$ is a somewhat involved function, chosen so that the line 
element (\ref{3.29}) takes the form

\begin{equation} ds_{\lambda}^2 = \frac{1}{4}\left( - \lambda^2
(d\psi - \cosh{\theta}d\phi )^2 + d\theta^2 + 
\sinh^2{\theta}d\phi^2 \right) \ . \end{equation}
 
\noindent An obvious advantage is that we recover the squashed 3-sphere, 
in Euler coordinates, through the replacement $\theta \rightarrow i\theta$. 
In the anti-de Sitter case these coordinates are related to the embedding 
coordinates by 

\begin{equation} X = \sinh{\frac{\theta}{2}}\cos{\frac{\psi + \phi}{2}} 
\hspace{7mm} Y = \sinh{\frac{\theta}{2}}\sin{\frac{\psi + \phi}{2}} 
\ \end{equation}

\begin{equation} U = \cosh{\frac{\theta}{2}}\cos{\frac{\psi - \phi}{2}} 
\hspace{8mm} Y = \cosh{\frac{\theta}{2}}\sin{\frac{\psi - \phi}{2}} \ . 
\end{equation}

\noindent Note that $\theta$ is a radial coordinate, $\theta > 0$. The 
manifest symmetries are generated by 

\begin{equation} \tilde{J}_0 = 2\partial_\psi \ , \hspace{8mm} 
J_0 = - 2\partial_\phi \ . \end{equation}

\end{document}